\documentclass[pre,a4paper,superscriptaddress,twocolumn,showpacs,amsmath,amssymb,floatfix]{revtex4}

\usepackage{graphicx}
\usepackage{amssymb}
\usepackage{float}

\input{epsf}

\begin{document}

\title{Anderson transition for Google matrix eigenstates}

\author{Oleg V.Zhirov}
\affiliation{\mbox{Budker Institute of Nuclear Physics, 630090 Novosibirsk, Russia}}
\affiliation{\mbox{Novosibirsk State University, 630090 Novosibirsk, Russia}}
\author{Dima L. Shepelyansky}
%\homepage[]{http://www.quantware.ups-tlse.fr}
\affiliation{\mbox{Laboratoire de Physique Th\'eorique du CNRS, IRSAMC, 
Universit\'e de Toulouse, UPS, 31062 Toulouse, France}}

\date{February 2, 2015}
%\date{\today}

\begin{abstract}
We introduce a number of random matrix models
describing the Google matrix $G$ of directed networks.
The properties of their spectra and eigenstates
are analyzed by numerical matrix diagonalization.
We show that for certain models it is possible to
have an algebraic decay of PageRank vector
with the exponent similar to real directed networks.
At the same time the spectrum has no spectral gap and a 
broad distribution of eigenvalues in the complex plain.
The eigenstates of $G$ are characterized by the Anderson transition
from localized to delocalized states and a mobility edge curve
in the complex plane of eigenvalues.
\end{abstract}

\pacs{89.75.Hc, 89.20.Hh, 89.75.Fb}
%89.20.Hh       World Wide Web, Internet
%89.75.Hc       Networks and genealogical trees 
%05.40.Fb       Random walks and Levy flights
%72.15.Rn Localization effects (Anderson or weak localization) 
%89.75.Fb Structures and organization in complex systems

\maketitle

\section{Introduction}
%\subsection{subsection}% Maybe ...
The phenomenon of Anderson localization \cite{anderson1958}
appears in a variety of quantum physical systems 
including electron transport in disordered solids and waves in random media
(see e.g. \cite{gilles,mirlin}). It is usually analyzed in the frame of
Hermitian or unitary matrices. Recently, the localization properties  of
nonunitary complex matrices has been analyzed 
for Euclidean matrices \cite{skipetrov2013} in relation to
light and wave localization \cite{skipetrov2014}.

In this work we analyze the possibilities of Anderson like localization
and delocalization 
for the matrices belonging to the class of Markov chains and Google matrix $G$
\cite{brin,meyer}. Such matrices have
real nonnegative elements with the sum of elements
in each column being equal to unity.
For a directed network one first defines an adjacency matrix
 $A_{ij}$ which has element $1$ if a node  $j$ have a link pointing to
node $i$ and zero otherwise.
The columns with only zero elements ({\em dangling nodes})
are replaced by columns with $1/N$ where $N$ is the matrix size.
The elements of other columns are renormalized
in such a way that their sum becomes equal to unity
($\sum_i S_{ij}=1$, $S_{ij}=A_{ij}/\sum_i A_{ij}$).
Thus we obtain the matrix $S_{ij}$ of Markov transitions.
Then the Google matrix $G$ of the network takes the form
\cite{brin,meyer}:
\begin{equation}
   G_{ij} = \alpha  S_{ij} + (1-\alpha)/N \;\; .
\label{eq1} 
\end{equation} 
Here, the damping factor $\alpha$ is taken in the range
$0<\alpha \leq 1$. In the context of the World Wide Web (WWW)
the term $(1-\alpha)$ describes for a random surfer
a probability to jump on any node of the network.
The above construction of $G$ has been proposed by Brin and Page \cite{brin}
to describe the structure of the World Wide Web (WWW). 
For the WWW it is assumed that the Google search engine uses 
$\alpha \approx 0.85$ \cite{meyer}. We can also consider 
a generalized case of weighted
Markov transitions $S_{ij}$ corresponding to real positive
elements of $A_{ij}$ like happens for the world trade
network (see e.g. \cite{wtrade}).

The matrix $G$ belongs to the class of Perron-Frobenius 
operators,
its largest eigenvalue 
is $\lambda = 1$ and other eigenvalues have 
$|\lambda| \le \alpha$ \cite{mbrin,meyer}. 
The right eigenvector at $\lambda = 1$, which is called the PageRank ($GP=P$), 
has real nonnegative elements $P(i)$
and gives a probability $P(i)$ to find a random surfer at site $i$. 
It is possible to rank all nodes
in a decreasing order of PageRank probability $P(K(i))$
so that the PageRank index $K(i)$ counts all $N$ nodes
$i$ according their ranking, placing 
the most popular nodes
at the top values $K=1,2,3 ...$.
Usually for many real directed networks the distributions of number of 
ingoing and outgoing links
are described by a power law 
(see e.g. \cite{dorogovtsev}), generating an average 
approximately  algebraic decay of PageRank probability $P(K) \propto 1/K^\beta$
with $\beta \approx 0.9$. Some  examples of directed networks
can be found in \cite{gmatrixrmp}. 

It is important to note that matrices of Google class practically have not been 
studied in physical systems even if they naturally appear in the frame of
Ulam networks generated by the Ulam method for 
dynamical maps in a coarse-grained phase space
(see e.g. \cite{zhirov2010pre,ermann2010epjb,frahm2010epjb}).

Therefore, it is interesting to see if the phenomena of 
Anderson localization and Anderson delocalization transition
can appear for Google matrices. 
Certain indications on a possible Anderson transition for
the Ulam networks, built from dissipative maps, 
have been reported in \cite{zhirov2010pre}
with more detailed discussions presented in \cite{gmatrixrmp}. 
Thus, it would be useful to find random matrix models
which are able to reproduce typical properties of spectrum and PageRank decay
in real directed networks. However, the results presented in \cite{physrev}
show that the full matrix $G$ with  random matrix elements 
have an unrealistic spectrum and hence other random matrix models of
$G$ should be developed. The models discussed in \cite{giraudpre}
give certain indications of delocalization of eigenstates of $G$
but the spectrum of $G$ in these models has a large gap and is very far from 
the spectra of real directed networks.
With this aim we describe below a number of 
random Google matrix models and analyze the properties of their spectra 
and eigenstates. We use certain spectral
properties of small size {\it orthostochastic}  matrices
with $N=3,4$ established in \cite{karol}.

\section{Random matrix models of G}

We start from a description of various random matrix models of 
Google matrix $G$ presenting the results of their spectral properties in
next Section. 

\subsection{Model RMZ3: random three-diagonal blocks} 
%s2.1

Following \cite{karol} we consider \textit{orthostochastic} matrix blocks 
$B_{ij}$ of size  $M \times M = 4 \times 4$. The \textit{orthostochastic} property
means that $B_{ij}={O_{ij}}^2$, where an orthogonal matrix $O$
has random matrix elements obtained via random rotations.
Since $O$ is  an orthogonal matrix the matrix $B$ is
bistochastic with $\sum_i B_{ij}= \sum_j B_{ij} = 1$ \cite{karol}.
The main reason to use such blocks $B$ is a similarity of
complex spectrum of random matrix ensemble of $B$
with the spectrum of $G$ of university networks of Cambridge and Oxford,
as discussed in \cite{gmatrixrmp}. 
The size $4 \times 4$ can be considered as preferential
random links between a group of $4$ friends.
However, a weak point of 
the random ensemble of $B$ \cite{karol} is a small matrix size $N=4$, while
for the above universities we have $N \approx 2 \times 10^5$. 

\begin{figure}
  \includegraphics[width=0.9\columnwidth]{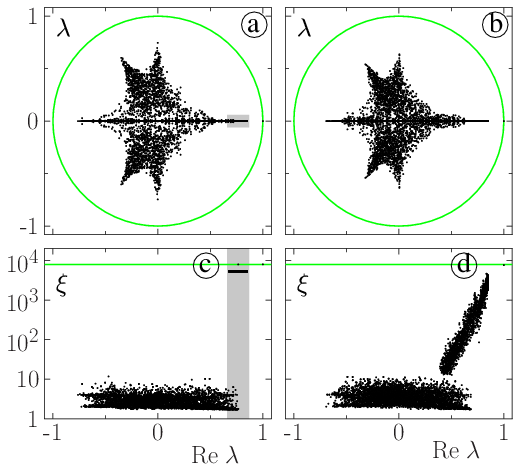}%
  \caption{\label{fig1}
     Google matrix eigenvalues $\lambda$ (a,b), and  IPR 
     $\xi$ of eigenvectors  as a function of ${\rm Re} \lambda$
     (c,d). Panels show data for RMZ3 model (Sec.2.1) at fixed 
     amplitudes $\varepsilon_i= 0.5$ (a,c) and 
     for random amplitudes 
      $ 0.15 \leq \varepsilon_i \leq 0.3$ (b,d).     
    The green circle shows $|\lambda|=1$ (a,b);
    the green horizontal line shows maximal possible $\xi=N$ (c,d);
    the gray band in (a,c)  highlights specific states (see text). 
    Here the total number of nodes is  $N$=8000. 
 }
\end{figure}

To go to large values of $N$ in matrix $S_{ij}$ 
we construct the {\bf Random Matrix model $Z3$ (RMZ3)}
as follows:
we place blocks $B$ of size $M=4$
on the main diagonal with weights $(1-\varepsilon_i)$
and on two adjacent 
upper and lower diagonals with weight
$\varepsilon_i/2$, where $\varepsilon_i$ $(i=1,\ldots,N/M)$ are 
random numbers uniformly distributed in some interval 
$(\varepsilon_{min},\varepsilon_{max})$;
each block represents a random realization of $B$;
then the matrix $G$ of total size $N$ is built from $S$
via the equation (\ref{eq1}).
Here we consider two cases with a constant
$\varepsilon_i=0.5$ and the interval range 
$0.15 \leq \varepsilon_i \leq 0.3$ (see Fig.~\ref{fig1}).
Obviously, by construction the final matrix belongs to the Google matrix class.
We use notations $S_Z$ and $G_Z$ for the matrices $S$ and $G$ of this model.

\subsection{Model RMZ3S: RMZ3 with shortcuts} 
%s2.2

\begin{figure}
  \includegraphics[width=0.9\columnwidth]{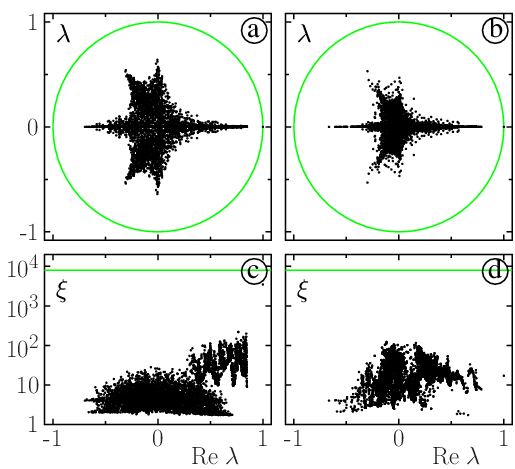}%
  \caption{\label{fig2} 
    Same as in Fig.~\ref{fig1} for RMZ3S model (Sec.2.2)
    with $0.15 \leq \varepsilon_i \leq 0.3$, shortcut amplitude
    $\varepsilon_s=0.3$, $\delta=0.1$ (a,c) and $\delta=1$ (b,d).
    Here $N=8000$. 
 }
\end{figure}

\begin{figure}
  \includegraphics[width=0.9\columnwidth]{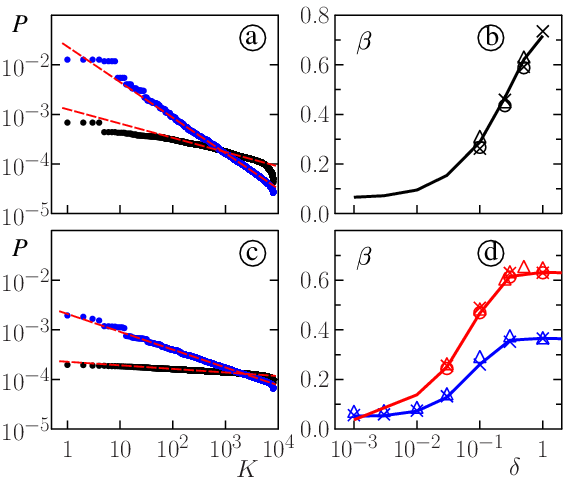}%
  \caption{\label{fig3}
   Dependence of $P(K)$ for models RMZ3S  (Sec.2.2) in panels (a,b)
   and RMZ3F (Sec.2.3) in panels (c,d); here $0.15 \leq \varepsilon_i \leq 0.3$.
   In panel (a) we have 
   $\delta=0.1$ (black symbols) and $\delta=1$ (blue symbols) at $N=8000$; 
   the fitted algebraic 
   dependence is shown by straight dashed lines with
   parameters:  $a=-6.67$, $\beta =0.288$ at $\delta=0.1$
   and  $a=-3.76$, $\beta = 0.71$ at $\delta=1$; panel (b) 
   shows the dependence of $\beta$ on $\delta$
   with the full curve for $N=8000$ and triangles, crosses and circles for
   $N=2000, 4000, 16000$ respectively; the amplitude of shortcut
   elements is $\varepsilon_s=0.3$. In panel (c)
   we have $\delta = 0.01$ (black symbols) and $\delta = 3$ (blue symbols) at 
   $\mu = 0.1$ and $N=8000$;
   the fitted algebraic 
   dependence is shown by straight dashed lines with
   parameters: $a=-8.39$, $\beta =0.072$ at $\delta=0.01$
   and $a=-6.17$, $\beta =0.36$ at $\delta=3$;  panel
   (d) shows the dependence of $\beta$ on $\delta$
    for $\mu =0.1$ (blue) and $\mu =0.3$ (red)
     with the full curves for $N=8000$ and triangles, crosses and circles for
   $N=2000, 4000, 16000$ respectively. 
 }
\end{figure}

The model {\bf RMZ3S} is obtained from RMZ3 by adding 
shortcut links between blocks $B$ in the upper triangle of the whole matrix $S$,
the  blocks of shortcut links are placed randomly in this part of $S$.
The amplitude of transitions from one block to another block
(outside of three-diagonal blocks of RMZ3) is taken at some
fixed value $\varepsilon_s$. The shortcut blocks are randomly and uniformly
distributed over the upper triangle of the whole matrix.
{\it After} adding the shortcut blocks the columns 
 affected by shortcut blocks are renormalized to unity.
In this way the obtained matrix $S$ again belongs to the Google matrix class.
The blocks of shortcuts are placed randomly in the upper triangle of matrix $S$,
their number $N_s$ is determined by the parameter
$\delta = 4 N_s/(3N)$. In fact $\delta$ gives the ratio of shortcut blocks 
to the number of blocks $3N/4$ in the model RMZ3.
Again each block $B$ in the main three-diagonal part of RMZ3
and blocks at shortcut positions are taken as random and independent realizations
for each block.
We note that the shortcuts between single nodes have been used for 
studies of quantum chaos and Anderson transition in the 
small world 
Anderson model (see \cite{smallworld1,smallworld2,gmatrixrmp}).
The results for RMZ3S model are shown in Figs.~\ref{fig2},~\ref{fig3}.

\subsection{Model RMZ3F: RMZ3 plus triangular matrix}
%s2.3

\begin{figure}
  \includegraphics[width=0.9\columnwidth]{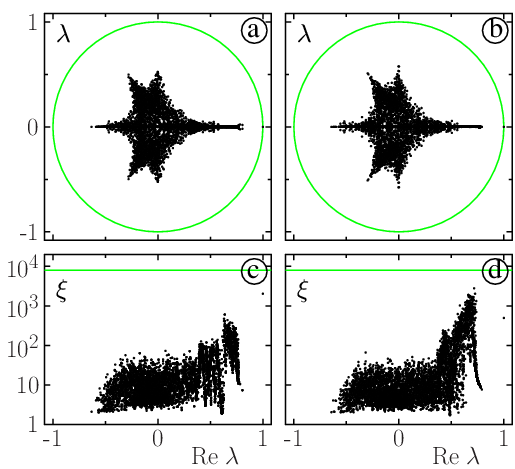}%
  \caption{\label{fig4}
    Spectrum (a,b) and IPR $\xi$ dependence of ${\rm Re} \lambda$
    for the model RMZ3F (Sec.2.3) at $\delta = 0.1$ (a,c) and
    $\delta =3$ (b,d); here $0.15 \leq \varepsilon_i \leq 0.3$,
    $\mu =0.1$, $N=8000$; circle and horizontal lines are as in Fig.~\ref{fig1}.
 }
\end{figure}

\begin{figure}[!h]
  \includegraphics[width=0.9\columnwidth]{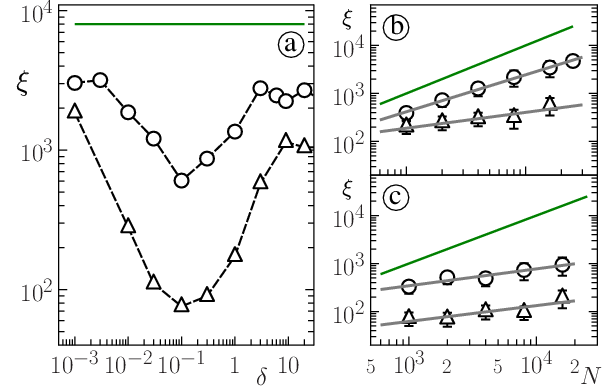}%
  \caption{\label{fig5}
   Panel (a) shows dependence of maximal IPR $\xi$ 
   (for states with $|\lambda|<1$) on
   parameter $\delta$ for the model RMZ3F (Sec.2.3)
   at $N=8000$. Dependence of maximal IPR $\xi$ on $N$
   is shown in panels (b) at $\delta=3$ and (c) at
   $\delta=0.1$; error bars show statistical error,
   if it is larger than symbol size, 
   obtained from $N_r$ disorder realizations. 
   We use $N_r=11$ at $N=2000$,
   $N_r=8$ at $N=4000$, $N_r=4$ at $N=8000$,
   $N_r=3$ at $N=16000$.
   In all panels
    $\mu=0.1$ (circles) and $\mu=0.3$ (triangles),
     $0.15 \leq \varepsilon_i \leq 0.3$;
   the straight green lines show dependence $\xi=N$; 
   the straight gray lines
   in (b,c) show the fitted dependence (see text). 
 }
\end{figure}

The results obtained in \cite{physrev,integers} show that
a triangular matrix of Google matrix class
has a tendency to have a realistic PageRank probability
decay with $P \propto 1/K$ and have some eigenvalues of finite amplitudes
$|\lambda|$. Due to these indications we construct
a matrix $S_F$ in the following way: $N_u$
random numbers $f_i$ from the interval $(0,1)$
are placed on random positions of the upper triangle of 
matrix of size $N$, then all columns are renormalized to unity
and columns with  all zero elements are replaced by
columns with all elements $1/N$. Then we construct the 
matrix $G$ of the {\bf model RMZF} as: 
\begin{equation}
   S_{ZF}=(1-\mu) S_Z + \mu S_F \, , \; G_{ZF}=\alpha S_{ZF}+ (1-\alpha)/N \; .
\label{eq2} 
\end{equation} 
Here $\mu$ determines a measure of contribution of $S_F$
with $0 < \mu <1$. The number of nonzero random elements $N_u$
is given by parameter $\delta=N_u/(12 N)$.
The results for the  RMZF model are shown in Figs.~\ref{fig3},\ref{fig4},~\ref{fig5}.

\subsection{Anderson models AD2 and AD3 for G matrix}
%s2.4

We use the usual Anderson model \cite{anderson1958,mirlin} 
with diagonal disorder terms $W_i$ and transitions $V$ to nearby sites on a lattice 
in dimension $d$: 
\begin{equation}
   W_{\bf i} \psi_{\bf i} + V \psi_{{\bf i+1}}+ V \psi_{{\bf i-1}}= \lambda \psi_{\bf i} \; ,
\label{eq3} 
\end{equation}
where indexes in bold are vectors in $d$-dimensional space.
On the basis of (\ref{eq3}) we construct the matrices
$S$ and $G$. 

Thus we consider the dimensions $d=2,3$ corresponding to
square and cubic lattices. The matrix $S$ is constructed as follows:
each transition matrix element, corresponding to $V$ terms,
in the Anderson model in dimension $d$  (\ref{eq3})
is replaced by a random number $\varepsilon_i$ uniformly
distributed in the interval $[0,\varepsilon_{max}/2d]$,
the diagonal element $W_ {\bf i}$ is replaced by unity minus
the sum of all $\varepsilon_i$ over $2d$ nearby sites 
($1-\sum_{i=1}^{2d} \varepsilon_i$).
The asymmetric matrix $S$ constructed in this way belongs to
the Google matrix class. Thus we obtain the matrices
$S_{AD2}, G_{AD2}$ for the {\bf model AD2}
and $S_{AD3}, G_{AD3}$ for the {\bf model AD3}
for $d=2$ and $3$ respectively.
The results for these models are presented in 
Figs.~\ref{fig6},~\ref{fig7},~\ref{fig9}.

\subsection{Anderson models AD2S and AD3S with shortcuts}
%s2.5

By adding shortcut links between pairs of nodes 
randomly distributed 
in the upper triangle of matrix $S$ we obtain {\bf models AD2S and AD3S}
respectively from models AD2 and AD3. The number of 
shortcut elements in $S$ is taken to be $N_s=2 d N \delta$,
their amplitude is taken as 
$0 \leq \varepsilon_i \leq \varepsilon_s=\varepsilon_{max}/2$,
after adding shortcuts the columns with shortcut elements are 
renormalized to unity.
Thus the sum of elements in each column is equal to
unity and $S$ belongs to the Google matrix class.
We note the matrices of these models as
$S_{AD2S}, G_{AD2S}, S_{AD3S}, G_{AD3S}$
respectively for $d=2,3$.
The results for these models are presented in 
Figs.~\ref{fig8},~\ref{fig9},~\ref{fig10}.

\subsection{Anderson models AD2Z and AD2ZS with blocks and block shortcuts}
%s2.6

By replacing matrix elements in the model AD2 by blocks $B$ of size $4 \times 4$
(see Sec.2.1) we obtain the {\bf model AD2Z}. In a similar way 
for the model AD2S we obtain the {\bf model AD2ZS} with block shortcuts.
In this case we restrict our studies only for dimension $d=2$
since the matrix size becomes too large for $d=3$.
Amplitudes $\varepsilon_{max}$ and $\varepsilon_s$ 
are defined as for the models AD2 and AD2S.
Since the transitions are now given by blocks
then the parameter $\delta$ is now defined as
$N_s=2d(N/4)\delta$ with $d=2$. 
The results for models AD2Z and AD2ZS
are presented in Figs.~\ref{fig11},~\ref{fig12}.

\begin{figure}
  \includegraphics[width=0.9\columnwidth]{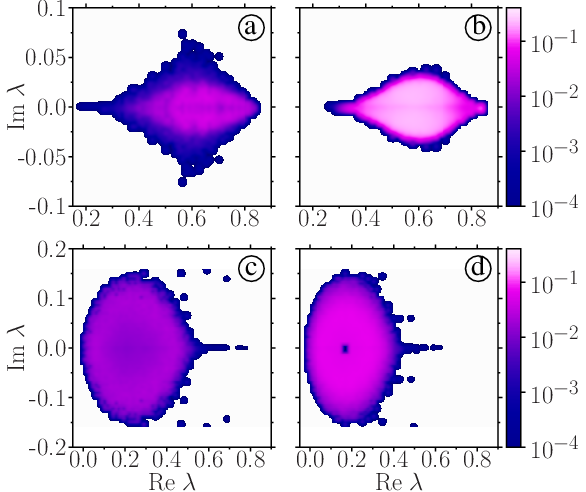}\\%
  \caption{\label{fig6}
  Distribution of IPR $\xi_i$ on $\lambda-$plane
  for the Anderson type models AD2  at $d=2$ (a)
  and AD3 at $d=3$ (b) (Sec.2.4) and
  the Anderson type models with shortcuts
    AD2S (c) and AD3S (d) at $\delta=2$  (Sec.2.5). Here
   $\varepsilon_{max} = 0.6$;  $N=130^2=16900$ for (a,c);
   $N=25^3=15625$ for (b,d) and  $\varepsilon_{max}=0.6$,
   $\varepsilon_s=\varepsilon_{max}/2=0.3$  
   for (c,d).
   Color bars show the ratio $\xi_i/N$
   (IPR values are averaged inside cells of 
   coarse-grained lattice $60 \times 60$).
 }
\end{figure}

\section{Spectral properties of G matrix models}
%s3

We use exact numerical diagonalization for
analysis of spectrum and eigenstates
of models of Sec.2. The matrix size $N$ 
is changed from a minimal $N=900$ up to maximal
$N=27000$.  For the description of the decay
of PageRank probability we use a fit
$\ln P = a - \beta \ln K$ 
which gives us the PageRank exponent of algebraic  $\beta$.
In all simulations we use $\alpha=0.85$. 
The right eigenstates $ {\psi_i}(j)$ of $G$ are determined by
\begin{equation}
   \sum_{j'=1}^N G_{jj'} {\psi_i}(j') = \lambda_i  {\psi_i}(j) \; .
\label{eq4} 
\end{equation} 
We characterize $ {\psi_i}(j)$ 
by the Inverse Participation Ratio (IPR)
$\xi_i=(\sum_j |\psi_i(j)|^2)^2/\sum_j |\psi_i(j)|^4$. This quantity 
is broadly used
in the studies of Anderson localization \cite{mirlin}
and determines the number of sites effectively populated by an eigenstate.
The value of $\xi$ is independent of normalization.
We use normalization $\sum_i P(i)=1$ for the PageRank eigenstate at $\lambda=1$.
For each eigenvector $\psi_i(j)$ we can order all nodes 
in a monotonically decreasing
order of $|\psi_i(j)|$ thus obtaining the local rank index $K$
for a given  $\psi_i(j)$. Such a ranking was used in 
\cite{ukuniv,wikispectrum}.
Below we describe the results for models of Sec.2.

\subsection{Results for RMZ3 model}
%s3.1

For the model RMZ3 at $\varepsilon_i = const =0.5$
the spectrum is shown in Fig.~\ref{fig1}a. We see that it has a form of 
$6-$rays star typical for the directed networks
studied in \cite{ukuniv,wikispectrum,gmatrixrmp}.
The size of the star is slightly reduced 
since all $\lambda_i(\alpha) \rightarrow \alpha \lambda_i(\alpha=1)$ 
for $\alpha < 1$, except $\lambda=1$ \cite{meyer}. There is also
addition reduction of $|\lambda_i|$ due to 
finite coupling terms   $\varepsilon_i >0$ but this reduction
is moderate and the spectrum of $G_Z$ is close to the spectrum
of independent $4 \times 4$ blocks found in \cite{karol}. 
Thus RMZ3 model captures a part of real properties of directed networks.

An interesting property of eigenstates becomes visible from
the dependence of $\xi$ on ${\rm Re} \lambda$ shown in Fig.~\ref{fig1}c
at  $\varepsilon_i =0.5$. Many eigenstates have relatively small
$\xi < 10$ which remain bounded with the increase of $N$ up to
$N =16000$ (data not shown). However, there is a group of states
(gray band) with $\xi \sim N$  growing linearly with $N$
(data not shown). These are delocalized states.
Their origin becomes clear from the following consideration.
We can use the anzats in which the elements of $\psi(j)$
are constant inside a given block $B_m$ with a values $\varphi_m$. 
Then Eq.(\ref{eq4}) takes the form
\begin{equation}
   (1-\varepsilon)\varphi_m + \varepsilon (\varphi_{m+1}+ \varphi_{m-1})/2= \lambda \varphi_m \; ,
\label{eq5} 
\end{equation}
since the matrix $G$ is bistochastic with sum of elements in rows being
unity since  $\varepsilon_i = const$. The spectrum $\lambda$ in (\ref{eq5}) is real.
Thus we obtain in (\ref{eq5}) the Bloch
equation with plane wave delocalized solutions well known for crystals \cite{gilles,mirlin}.
These solutions belong to the gray band part of the spectrum in Fig.~\ref{fig1}a. 
Another part of the spectrum corresponds to such $\psi(j)$ which have 
different values on a scale of one block $B$. 

The case with different $\varepsilon_m$ 
(e.g. $0.15 \leq \varepsilon_m \leq 0.3$ in Fig.~\ref{fig1}b,d)
we can use the same anzats for the left vector $\varphi_m$
that leads to the eigenvalue equation:
\begin{equation}
   (1-\varepsilon_m)\varphi_m + \varepsilon_m (\varphi_{m+1}+ \varphi_{m-1})/2= \lambda \varphi_m .
\label{eq6} 
\end{equation}
Such a problem corresponds to the case of off-diagonal disorder 
in the $1d$ Anderson model where the localization length, and hence IPR, is diverging 
at the center of the band \cite{gilles,mirlin}. 
The spectrum $\lambda$ in (\ref{eq6}) is real.
A  similar problem is known as the Sinai walk \cite{sinai}
where transition probabilities on a Markov chain are fluctuating.
This model has been studied extensively 
(see e.g. \cite{ledoussal} and Refs. therein).

The spectrum $\lambda$ in (\ref{eq6}), corresponding to this anzats,
is the same for the right eigenvectors \cite{meyer}. 
The right eigenvectors are different 
from the left ones but have a similar structure on average.
The IPR values, shown in Fig.~\ref{fig1}d are significantly reduced,
comparing to the case $\varepsilon_m=const$, 
except those with $\lambda$ close to unity. When $N$ is increasing
we find that IPR is growing only for $\lambda \rightarrow 1$
while for $|\lambda| <1$ IPR values remains finite.
This corresponds to the known results for 
the Anderson model with off-diagonal disorder.
Other eigenstates for which $\psi_i$ is not constant inside $B$ blocks
correspond to the eigenstates with rather small IPR values $\xi \sim 10$.

Even if the spectrum and eigenstates have interesting properties in the two above cases
of model RMZ3 there is a weak point here: 
the PageRank probability $P$ in these cases is flat 
being practically independent of $K$ and $\xi \sim N$. 
Thus the situation is very different from the real directed networks
with $\beta \approx 1$ (see e.g. \cite{meyer,gmatrixrmp}). 
This happens due to a space homogeneous structure of the matrix $G$
(a part of fluctuations) and thus there is no leading node with 
a large number of links. Due to that 
we try to introduce shortcut links as described in the next Sec.3.2.

\subsection{Properties of RMZ3S model}
%s3.2

The spectrum and IPR dependence for RMZ3S model with shortcuts
are shown in Fig.~\ref{fig2} for two typical values
of parameter $\delta$. We see that at small values of $\delta $ (e.g. $\delta=0.1$)
the spectrum structure is practically the same as for RMZ3 model.
However, for larger values ( $\delta=1$) the size of the spectrum star 
is decreasing. The values of IPR are significantly reduced
at finite values of $\delta$ and our data show that the maximal
$\xi$ values remain less than $\xi = 200$ even for the largest 
size $N=16000$ for $0.1 \leq \delta \leq 1$ for 
all $|\lambda|<1$ (data not shown). Thus in this model
all eigenstates remain localized.

Even if all states are localized the decay of PageRank is more close 
to the case of real directed networks.
Indeed, the data of Fig.~\ref{fig3}a,b show that $P(K)$
have approximately algebraic decay with PageRank index.
The fit allows to determine the PageRank exponent $\beta$
which is small at $\delta \sim 0.1$ and 
is growing with increase of $\delta$ reaching values $\beta \approx 0.75$
at $\delta =1$. 
It is important to note that $\beta$ is independent of $N$
at large $N$ values.
Thus the homogeneous random elements in the upper triangle
of $S$ matrix allow to obtain $\beta$ close to unity
at large $\delta$. 
Indeed, in the limit of rather large $\delta$
we come to the case of triangular matrix $S$ studied in \cite{physrev}
(and also in \cite{integers}) where one obtains 
an approximate  decay $P \propto 1/K$.
Indeed, at large $\delta$ a sum of elements in a row of $G$
drops approximately as $1/K$ (where $K$ is a row index) leading to $P \propto 1/K$.
Indeed, we can say that $P(K) \approx \sum_j G_{Kj} e_j \sim 1/K$,
where $e_j=1/N$ is a homogeneous initial vector,
considering this as one iteration of the PageRank algorithm \cite{meyer}. 
We note that for the
PageRank vector we have $\xi \sim N$ for  $\beta < \beta_c=1/4$.

Thus the model RMZ3S has a reasonable spectrum structure
and an algebraic PageRank probability decay.
But all eigenstates with $|\lambda|<1$ remain localized.
Thus we go to the analysis of RMZ3F model.

\subsection{Results for RMZ3F model}
%s3.3

The spectrum and IPR values for the RMZ3F model
are presented in Fig.~\ref{fig4}. We see that the star spectrum
structure is preserved but IPR values are increased
in a vicinity ${\rm Re} \lambda \approx \alpha$.
The examples of $P(K)$ and $\beta(\delta)$ dependencies 
are shown in Fig.~\ref{fig3}c,d. 
It is important to note that $\beta$ is independent of $N$
at large $N$ values.
Qualitatively, the situation
is similar to the model RMZ3S but the effect of $\delta$ on localization
properties of $\xi$ is more complicated. 

Indeed, it is well seen in Fig.~\ref{fig5}a
that the maximal IPR values (excluding PageRank vectors)
are at first reduced with an increase of $\delta$ from 
$10^{-3}$ up to $0.1$ but they are increased 
when $\delta$ goes from $0.1$ to $10$.
The dependence of maximal $\xi$ on $N$
at $\delta=0.1; 3$ is shown in Fig.~\ref{fig5}b,c
for $\mu=0.1; 0.3$. We fit this dependence by
a power law $\xi \propto N^\nu$ and obtain
for $\mu=0.1$: $\nu=0.352$ (at $\delta=0.1$)
and $\nu=0.770$  (at $\delta=3$);
for $\mu=0.3$:  $\nu=0.33$ (at $\delta=0.1$ and $3$).
These results show that there are certain states 
(except PageRank) that become delocalized in the limit
of large matrix size. In a certain sense, 
for the dependence $\xi(N)$ we have a certain
similarity with the results obtained in 
\cite{giraudpre} where a sub-polynomial
growth of $\xi$ with $N$ has been found
for randomized university networks and 
preferential attachment models.
However, for the RZ3F model the spectrum 
has no large gap and is more similar to the 
real directed networks. 

%OLD TEXT:
%Dependence of IPR on $\delta$ and $N$  (Fig.\ref{fig5}). 
%The latter dependence is fitted
%by formula $\ln \xi(N)= a+b \ln N$ (gray lines). For $\mu=0.1$: 
%$a=3.42$ and $b=0.352$ (at $\delta=0.1$), and 
%$a=0.711$ and $b=0.770$   (at $\delta=3$). 
%For $\mu=0.3$: $a=1.85$ and $b=0.33$ 
%(at $\delta=0.1$) and $a=  2.96 $ and $b= 0.33$ (at $\delta=3$) (Fig.\ref{fig5}):
 
The investigations of RMZ3F model at larger 
sizes (e.g. with the help of the Arnoldi method \cite{gmatrixrmp,ukuniv})
can provide a more firm conclusion about the delocalization properties 
of eigenstates in this model.

\subsection{Properties of AD2 and AD3 models}
%s3.4

\begin{figure}
  \includegraphics[width=0.9\columnwidth]{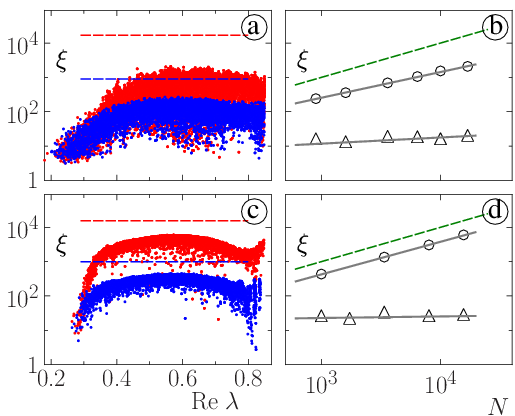}%
  \caption{\label{fig7}
   Dependence of $\xi$ on ${\rm Re} \lambda$ (a,c)
   and $\xi$ on $N$ (b,d) 
   for the models AD2 (a,b) and AD3 (b,d) (see Sec.2.4).
   For AD2: panel (a) is for  $N=900$ (blue, $N_r=10$ realizations)
   and $N=16900$ (red, $N_r=1$);  
   panel (b) shows dependence  $\xi(N)$ with fits $\xi \propto N^\nu$ for 
   eigenstates at the spectrum edge with  ${\rm Re} \lambda=0.23-0.25$ (triangles,
    $\nu=0.18$) and for maximal $\xi$
   (circles,    $\nu=0.75$).
   For AD3: panel (c) is for  $N=1000$ (blue, $N_r=10$ realizations)
   and $N=15625$  (red, $N_r=1$); panel (d)  shows dependence  $\xi(N)$ for 
   eigenstates at the spectrum edge  with  ${\rm Re} \lambda=0.23-0.25$ (triangles,
     $\nu=0.05$) for maximal $\xi$
   (circles,   $\nu=0.95$). Here $\varepsilon_{max} = 0.6$. 
   The fits are shown by gray lines, green (b,d) and blue, 
   red (c,d) dashed lines show
   dependence $\xi=N$. For panels (b,d) the number of realizations
   changes from $N_r=10$ to $3$ when $N$ changes from minimal to maximal value. 
 }
 \end{figure}

The spectra of AD2 and AD3 models are shown in
Fig.~\ref{fig6}a,b with color plot of IPR values.
We see that there are rather large values of $\xi$
indicating existence of delocalized eigenstates.
Indeed, a more detailed analysis
presented in Fig.~\ref{fig7}
shows that for the states of the spectral range
${\rm Re} \lambda > 0.25$ IPRs are growing with $N$
clearly demonstrating delocalization.
Indeed, for maximal  $\xi$ from this range 
(excluding PageRank)
we find $\nu =0.75$ at $d=2$
and $\nu =0.95$ at   $d=3$. At the same time 
in a vicinity of the spectrum edge
${\rm Re} \lambda < 0.25$ we have 
 $\nu =0.18; 0.05$ for $d=2;3$
clearly showing that in this part of the spectrum 
the eigenstates are well localized. 
Indeed, for these localized states 
we have an exponential decay 
$\ln |\psi| \propto - K^{1/d}$ with the eigenstate
rank index $K$ (see Fig.~\ref{fig9}a,b). 
Such a decay also
appears for the localized states of the Anderson model
in dimension $d$.  

But for the majority of eigenstates we have
significant growth of $\xi$ with $N$
showing that these states are delocalized.
 Of course, the case of $d=2$ should be
studied in more detail since 
for the standard Anderson model at $d=2$ (\ref{eq3}) 
all eigenstates are exponentially localized \cite{mirlin}.
However, we have here non-Hermitian matrix
and for our knowledge there are no
rigorous results about localization
in such matrices in $d=2$.

Even if in AD2, AD3 models we find delocalization, the 
PageRank in  these models
is practically flat
due to absence of central node
(see Fig.~\ref{fig9}a,b). Another weak point 
of AD2, AD3 models is a relatively narrow distribution of
eigenvalues with $|{\rm Im} \lambda|<0.1$
and due to that we continue our analysis with the next model.

\subsection{Results for AD2S and AD3S models}
%s3.5

\begin{figure}
    \includegraphics[width=0.9\columnwidth]{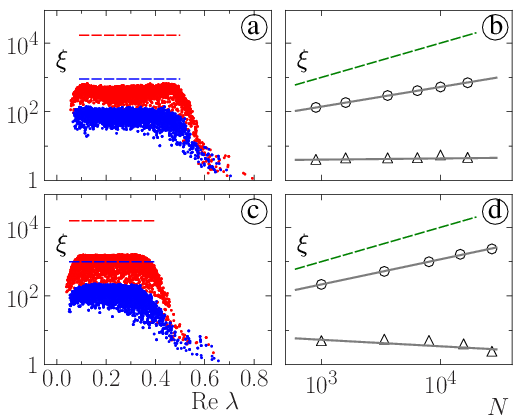}%
 \vglue 0.4cm
   \caption{\label{fig8}
     Same as in Fig.~\ref{fig7} but for the models 
     AD2S (a,b) and AD3S (c,d) (see Sec.2.5)
     at $\delta=2$; all parameters are as in Fig.~\ref{fig7}.
     The fits give: (b) $\nu= 0.04$ at the spectrum edge around
     ${\rm Re} \lambda \approx 0.6$ (triangles), $\nu=0.57$ for maximal $\xi$ (circles);
     (d)  $\nu= -0.19$ at the spectrum edge around
     ${\rm Re} \lambda \approx 0.6$ (triangles), $\nu=0.73$ for maximal $\xi$ (circles).
     Here $ \varepsilon_{max}=0.6$,
     $\varepsilon_s=\varepsilon_{max}/2=0.3$. For panel (b) [(d)] the number of realizations
     changes from $N_r=10$ to $3$ [$1$] when $N$ changes from $900$  to 16900 [$27000$]. 
  }
 \end{figure}

The spectra of AD2S, AD3S models are shown in Fig.~\ref{fig6}c,d.
We see that the additional terms in upper triangle of matrix $S$
produce a broadening of ${\rm Im} \lambda$ which however
still remains relatively narrow ($|{\rm Im} \lambda| < 0.2$).
The IPR values are growing with $N$ except of the eigenstates 
at the spectral edge ${\rm Re} \lambda \approx 0.6$ (see Fig.~\ref{fig8}).
For these localized states the exponent $\nu$ is practically zero
while for the maximal IPR (except PageRank)
we find rather large values of $\nu = 0.57$ at $d=2$,
$\nu = 0.73$ at $d=3$. Thus, in these models
we clearly have the Anderson type transition
from localized to delocalized eigenstates.

In analogy with the 3d Anderson model \cite{mirlin},
we make a conjecture that 
in models AD2, AD3, AD2S, AD3S
there is a
certain mobility edge curve
in the complex plane $\lambda$
which separates localized from delocalized states.
In a qualitative manner such a curve is
visible in Fig.~\ref{fig6}
as a border between blue color of localized
states with small $\xi$ and white color
of states with large $\xi$.
But definitely more detailed studies are required
for a more exact determination of such a mobility edge curve.

\begin{figure}
  \includegraphics[width=0.9\columnwidth]{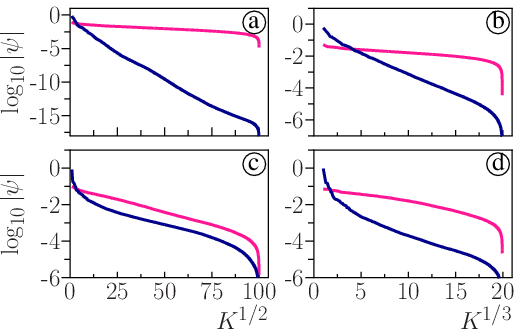}\\%
  \vglue 0.4cm
  \caption{\label{fig9} 
Dependence of eigenvector amplitudes $|\psi|$ on their rank index $K$
for models AD2 (a), AD3 (b) from (Sec.2.4)
and AD2S (c), AD3S (d) from (Sec.2.5).
Here $\delta=0$ for (a,b) and $\delta=2$ for (c,d);
$N=10^4$ for (a,c) and $N=20^3$ for (b,d).
We use $\varepsilon_{max} = 0.6$ and
$\varepsilon_s=0.3$ in (c,d).
Data show maximally delocalized (maximal $\xi$
corresponding to PageRank, magenta upper curve)
and maximally localized (smallest $\xi$, blue bottom curve) eigenstates.
 }
\end{figure}

\begin{figure}
   \includegraphics[width=0.9\columnwidth]{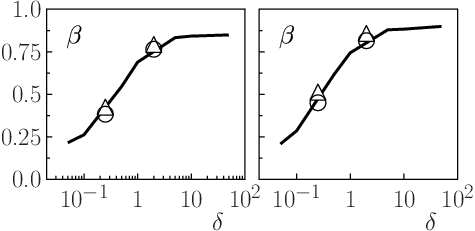}\\%
   \caption{\label{fig10}
  Dependence of the
  PageRank exponent $\beta$ on the parameter $\delta$
  for the models AD2S (left panel) and AD3S (right panel).
  Left panel: the solid curve shows data for 
  $N=80^2$, with triangles for  $N=40^2$
  and circles for  $N=130^2$. Right panel:
  the solid curve shows data for $N=20^3$, with
  triangles for  $N=10^3$ and circles for  $N=25^3$.
  Here $ \varepsilon_{max} = 0.6$ and $\varepsilon_s=0.3$
  }
 \end{figure}

Examples of PageRank probability decay are shown in Fig.~\ref{fig9}.
The new element, appearing in AD2S, AD3S models (comparing to AD2, AD3 cases),
is a dependence of the PageRank exponent $\beta$ on the parameter
$\delta$ as  shown in Fig.~\ref{fig10}.
These data demonstrate that $\beta  $ increases from $\beta \approx 0.2$
at $\delta =0.1$ up to $\beta \approx 0.9$ at $\delta =3$.
Thus AD2S,AD3S models have delocalized eigenstates and 
the PageRank exponent of real directed networks.
The only weak point is a narrow distribution of spectrum in ${\rm Im} \lambda$.
To improve this feature we study in next Section the models AD2Z, AD2ZS. 

\subsection{Results for AD2Z and AD3ZS models}
%s3.6
 
\begin{figure}
   \includegraphics[width=0.9\columnwidth]{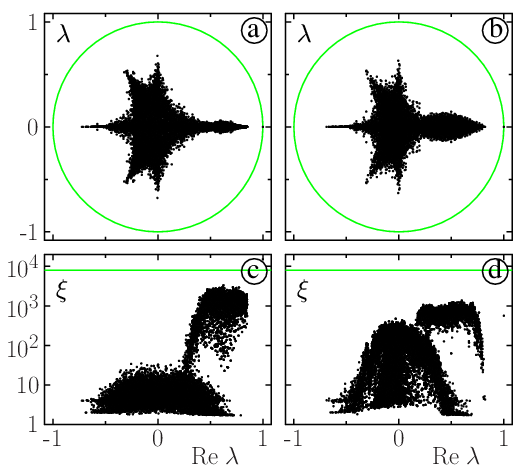}%
\vglue 0.4cm
   \caption{\label{fig11}
    Spectrum $\lambda$ (a,b) and IPR $\xi$ vs. ${\rm Re} \lambda$ (c,d) 
    for the models  AD2Z (a,c) and AD2ZS at $\delta=0.25$ (b.d)
    from Sec.2.6. Here $N=4 \times 70^2 = 19600$,
     $\varepsilon_{max} = 0.6$ and $\varepsilon_s=0.3$ in (c,d).
  }
 \end{figure}

The spectra of AS2Z, AD2ZS models are shown in Fig.~\ref{fig11}.
We see that the star structure appears due to 
introduction of blocks $4 \times 4$. 
The dependence of IPR $\xi$ on ${\rm Re} \lambda$
clearly shows the existence of two groups of states
with small $\xi < 100$, presumably for
localized phase, and large $\xi > 100$,
presumably for delocalized phase.

\begin{figure}
    \includegraphics[width=0.9\columnwidth]{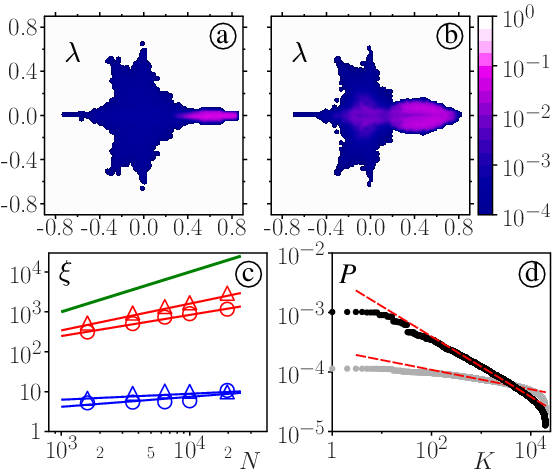}\\%
\vglue -0.4cm
    \caption{\label{fig12}
   Top panels show distribution of IPR $\xi$ values
   on $\lambda-$plane for models AD2Z (a) and
   AD2ZS at $\delta=0.25$ (b) of Sec.2.6 
   with parameters of Fig.~\ref{fig11};
   color bar gives the ratio $\xi/N$
   obtained from cells as in Fig.~\ref{fig6}.
   Panel (c): dependence of $\xi$ on $N$
   for AD2Z with triangles for states with $\lambda$
   located in the delocalized domain ${\rm Re} \lambda \in (0.3,0.85)$ 
   (red triangles, fit gives $\nu = 0.67$)
   and in the localized domain  ${\rm Re} \lambda < -0.5$ 
   (blue triangles, $\nu =0.15$);
    for AD2ZS at $\delta=0.25$ with circles for states with $\lambda$
   located in the delocalized domain ${\rm Re} \lambda \in (0.2,0.85)$ 
   (red circles, $\nu = 0.53$) 
   and in the quasi-localized domain  ${\rm Re} \lambda < -0.5$ 
   (blue circles, $\nu =0.25$);
   fits are shown by lines, green line shows $\xi=N$.
   Panel (d): dependence of PageRank probability $P$ on PageRank index $K$
   for models AD2Z (gray symbols) and AD2ZS at $\delta=0.25$ (black symbols);
   the fits for the range $K \in (100, 6000)$ are shown by dashed lines with
   $\beta=0.16$ (AD2Z) and $\beta=0.51$ (AD2ZS) for the parameters of panels (a,b).
   }
  \end{figure}

The distribution of $\xi$ on $\lambda-$plane is
shown in Fig.~\ref{fig12}a,b. Again we see signs
of the mobility edge curve separating localized
(blue) and delocalized (white) eigenstates.

The dependence of $\xi$ on $N$ is shown in 
Fig.~\ref{fig12}c. There are well localized states with 
$\xi$ practically independent of $N$
($\xi < 20$) and delocalized states for which
$\xi$ is growing with $N$
with a relatively large growth exponent
$\nu = 0.67$ at $\delta=0$ and
$\nu = 0.53$ at $\delta=0.25$.
This gives a strong argument 
for existence of the Anderson transition with
a mobility edge in a complex $\lambda-$plane
in these models.

The decay of PageRank probability is shown in Fig.~\ref{fig12}d:
at $\delta=0$ we have a flat $P(K)$ distribution 
with the exponent $\beta = 0.16$,
while at $\delta=0.25$ we find $\beta =0.51$
being close to the values found in real directed networks
(e.g. for the Twitter network $\beta \approx 0.54$ \cite{gmatrixrmp}).

Thus we can say that the model AD2ZS 
is the one being most close to real directed networks
with the number of interesting features:
algebraic decay of PageRank probability with the
exponent $\beta \approx 0.5$,
absence of spectral gap at $\alpha=1$,
a broad star like distribution 
of eigenvalues in the complex $\lambda-$plane,
existence of localized and delocalized eigenstates
of the Google matrix  with
strong indications on the Anderson transition
and the mobility curve in  $\lambda-$plane.

We expect that a similar model AD3ZS  constructed in dimension $d=3$
from the AD3S model will have even stronger delocalization properties.
 
\section{Discussion}

In this work we described various random matrix models of
the Google matrix of directed networks.
Our results show that for certain models (like AD2ZS)
we have an algebraic decay of PageRank probability with the exponent
$\beta \sim 0.5$, absence of spectral gap at $\alpha=1$,
existence of the Anderson transition and mobility edge 
in the complex $\lambda-$ plane. We think that the further analysis of the
models described here will allow to establish more close
links between the Anderson delocalization phenomenon in disordered solids
and   delocalization of eigenstates of the Google matrix of directed networks.

\section{Acknowledgments}
We thank L.Ermann and K.M.Frahm for useful discussions.
This research  is supported 
in part by the EC FET Open project 
``New tools and algorithms for directed network analysis''
(NADINE $No$ 288956). The research of OVZ was partially supported by the Ministry of
Education and Science of  Russian Federation.

\end{document}